# Magnetic nutation: transient separation of magnetization from its angular momentum

Anulekha De[1],[*] Julius Schlegel[2],[†] Akira Lentfert[1], Laura Scheuer[1], Benjamin Stadtmüller[1], Philipp Pirro[1], Georg von Freymann[1,3], Ulrich Nowak[2], and Martin Aeschlimann[1]

[1]Fachbereich Physik and Landesforschungszentrum OPTIMAS, Rheinland-Pfälzische Technische Universität Kaiserslautern-Landau, 67663 Kaiserslautern, Germany

[2] Fachbereich Physik, Universität Konstanz, 78457 Konstanz, Germany

[3]Fraunhofer Institute for Industrial Mathematics, ITWM 67663 Kaiserslautern, Germany

[*]Corresponding author: ade@rptu.de
[†]Corresponding author: julius.schlegel@uni-konstanz.de

For nearly 90 years, precession and relaxation processes have been thought to dominate magnetization dynamics. Only recently has it been considered that, on short time scales, an inertia-driven magnetization dynamics should become relevant, leading to additional nutation of the magnetization vector. Here, we trigger magnetic nutation via a sudden excitation of a thin $Ni_{80}Fe_{20}$ (Permalloy) film with an ultrashort optical pulse, that leads to an abrupt tilting of the effective field acting on the magnetic moments, separating the dynamics of the magnetization from that of its angular momentum. We investigate the resulting magnetization dynamics in the inertial regime experimentally by the time-resolved magneto optical Kerr effect. We find a characteristic oscillation in the Kerr signal in the range of ∼ 0.1 THz superimposed on the precessional oscillations with GHz frequencies. By comparison with atomistic spin dynamics simulations, we demonstrate that this observation cannot be explained by the well-known Landau-Lifshitz-Gilbert equation of motion but can be attributed to inertial contributions leading to nutation of the magnetization vector around its angular momentum. Hence, an optical and non-resonant excitation of inertial magnetization dynamics can trigger and control different magnetic processes, ranging from demagnetization via nutation to precession in a single device. These findings will have profound implications for the understanding of ultrafast spin dynamics and magnetization switching.

# INTRODUCTION:

The dynamics of magnetization in a magnetically ordered material is governed by precession [1]. Macroscopically, this precession leads to magnetic resonance phenomena that are used in a wide range of applications, from materials characterization to medical diagnostics. Microscopically, spin precession, or more precisely the precession of the spin's magnetic moment, explains the existence of spin waves and – in its quantized form – magnons as quasiparticles for the excitation of a magnetic ground state. Magnons are in the focus of contemporary spintronics [2, 3] with applications in sensor devices, data processing and energy efficient computing [4]. However, since spins are quantized angular momenta, there must be analogies to the mechanics of rotating bodies. If the rotation axis of a mechanical gyroscope is tilted away from the direction of the gravitational field by means of an external force, it will start to precess around the gravitational field, maintaining a constant angle with it. But there is a second type of dynamics: if one disturbs the precession such that the rotation axes of the gyroscope and its angular momentum are no longer aligned, an additional motion of the gyroscope around the angular momentum axis appears, known as the nutation [5, 6].

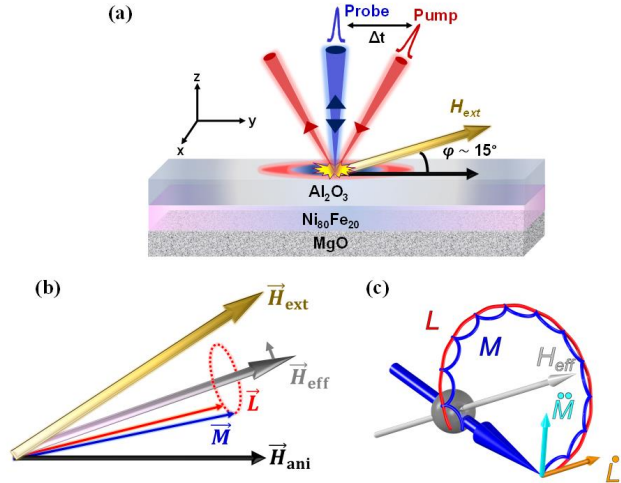

**FIG. 1.** (a) Schematic of the pump-probe experiment. (b) The pump pulse leads to a sudden reduction of the anisotropy field $\vec{H}_{ani}$ which, in turn, leads to a sudden tilting of the effective field $\vec{H}_{eff}$ so that the equilibrium direction for the magnetization vector $\vec{M}$ changes within some hundred fs. Consequently, the angular momentum $\vec{L}$ starts to precess. (c) Magnetic inertia causes the dynamics of the magnetization to be separated from the angular momentum dynamics. As the angular momentum starts to precess around the effective field, the magnetization accelerates toward the effective field direction, initiating nutation dynamics on top of precession.

In analogy to the magnetization dynamics, the axis of the gyroscope defines the magnetization. In principle, the orientation of the magnetization can deviate from the associated total angular momentum [7-11] (and in the following, we will not distinguish spin and orbital contributions to the total angular momentum of the electronic system). For a gyroscope, the nutation dynamics can be triggered by a sudden torque pulse. In this work, we demonstrate that – in complete analogy to the mechanical case – we can trigger magnetic nutation via a sudden excitation of the magnetic system of a thin $Ni_{80}Fe_{20}$ (Permalloy or Py) film with an ultrashort laser pulse (Fig 1a). As in the mechanical case, a "sudden excitation" is defined by its timescale being substantially shorter than that of the resulting nutation, so the laser pulse must be much shorter than one period of nutation. This ultrashort pulse then leads to an abrupt tilting of the effective field acting on the spins' magnetic moments, which initiates a precession of the angular momentum (Fig. 1(b)). Crucially, the magnetic inertia leads to a separation of the dynamics of the magnetization from that of the angular momentum of the spin system, initiating nutation. This effect is sketched in Fig. 1(c): In the absence of inertia, these dynamics would set in instantaneously with an abrupt change of the magnetic velocity, and magnetization and angular momentum would always be parallel. Considering inertia, the kinetic energy cannot increase instantaneously. Instead, as we will show later on, the magnetization starts to accelerate towards the effective field direction giving rise to nutation. The additional nutation is clearly a consequence of the sudden change of the effective field direction.

According to theoretical predictions [7-19] nutation could add a rich variety of effects to the magnetization dynamics, including new nutational resonances [8, 19], nutational spin waves [20-22], a shift of the known precessional resonances [19, 23], switching processes driven via a resonant excitation of nutation [24, 25] and even nutational auto-oscillations [26] appear to be possible. However, all these studies are purely theoretical and – despite its fundamental importance – the experimental investigation of nutation effects is still in its infancy. Only recently, the resonant excitation of inertial spin dynamics has been experimentally observed, where THz pump-optical probe measurements were used to trigger and detect the nutation resonance [27, 28]. However, our results go beyond the mere demonstration of the existence of magnetic nutation, as previously shown in [27, 28], but show that the separation of the angular momentum from its magnetic moment is an inherent phenomenon occurring during the relaxation of a non-equilibrium excitation of a magnetic system using ultrashort optical pulses. This "separation" induced by nutation means that the angular momentum and magnetic moment vectors are no longer parallel to each other. This leads to a qualitatively different behavior of the dynamics of magnetization ($\vec{M}$) and angular momentum ($\vec{L}$) - a nutating and precessing $\vec{M}$, but only a precessing and non-nutating $\vec{L}$. Moreover, we provide a detailed theoretical understanding of the observed nutation dynamics. Thus, our study allows us to control the integration of different processes of magnetization dynamics, ranging from ultrafast demagnetization via nutation to precession in a single experiment.

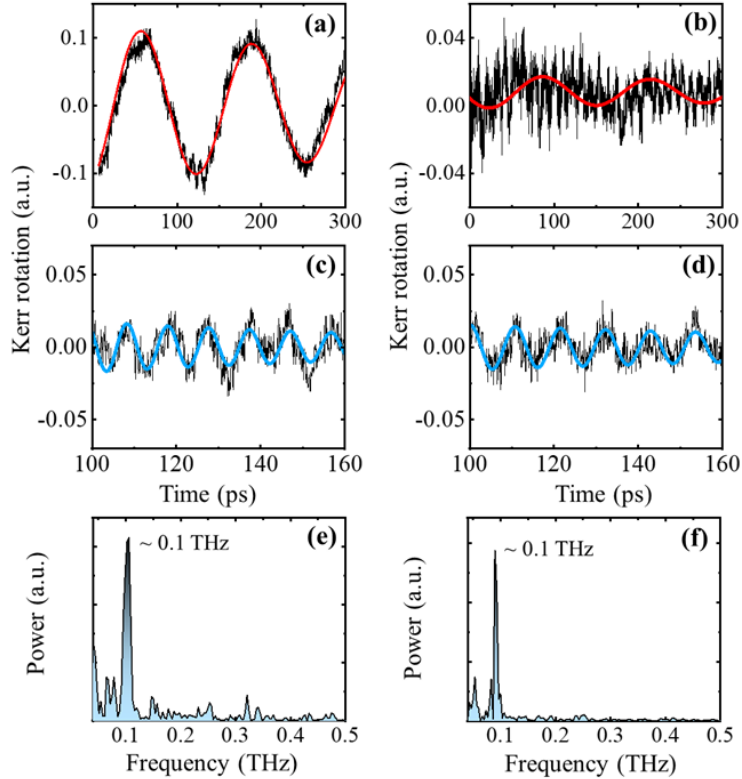

**FIG. 2.** Time-resolved Kerr rotation data measured up to 300 ps for (a) 5 nm and (b) 2.8 nm Py samples at $\mu_0 H_{ext} = 113$ mT and $F = 4.6$ mJ cm$^{-2}$. The black lines are the experimental data and the red lines represent the FMR background. Zoomed views of the background subtracted TR data of (c) 5 nm and (d) 2.8 nm Py samples for small time window (100 ps – 160 ps) showing nutational oscillations. The black lines are the experimental data and the blue lines are the sinusoidal fit to the data points. The FFT power spectra from the background subtracted entire TR data (upto 300 ps) of (e) 5 nm and (f) 2.8 nm Py sample showing a strong peak at ~ 0.1 THz. The $x$ axis is plotted from 30 GHz to 0.5 THz.

## II. RESULTS AND DISCUSSION:
### A. Experimental results:

Spin dynamics triggered by sudden excitation of a thin $Ni_{80}Fe_{20}$ film (thickness of 5 nm and 2.8 nm) with an ultrashort optical pulse was observed using a time-resolved magneto-optical Kerr effect (TR-MOKE) setup based on a two-color, non-collinear pump-probe technique (Fig 1(a) and the Supplemental Materials [29]). The samples are epitaxially grown on double-sided polished MgO substrates by molecular beam epitaxy (MBE) technique at room temperature and are capped with 3 nm $Al_2O_3$ layer. After the femtosecond (fs) pump pulse, the magnetization of the system is partially or completely lost within hundreds of fs, which is known as ultrafast demagnetization [30-32]. This is generally followed by a fast recovery of magnetization within sub-ps to a few ps and a slower recovery within hundreds of ps, known as fast and slow remagnetization. The slower recovery is accompanied by a precession of the magnetization [32-35]. On the shorter time scales of this slow recovery period (< 300 ps), we have observed that the usual precessional dynamics is enriched by an additional oscillation with higher frequency. To quantify the frequencies of these oscillations, we have recorded TR-MOKE data up to 300 ps with very high time step resolution (as shown in Fig. 2(a) and (b) for 5 nm and 2.8 nm Py samples, respectively measured at $\mu_0H_{ext}$ = 113 mT and $F$ = 4.6 mJ cm$^{-2}$) and subtracted the Ferromagnetic Resonance (FMR) background with the frequency obtained from the FFT of the precessional oscillations [32]. The precessional data up to longer time delays are provided in the Supplemental Materials [29]. The zoomed views of the background subtracted TR data for small time window showing nutational oscillations are depicted in Fig. 2(c) and (d). The TR-MOKE data up to 300 ps at $\mu_0H_{ext}$ = 96 mT and $F$ = 4.6 mJ cm$^{-2}$ are provided in the Supplemental Materials [29]. The FFT power spectra obtained from the FMR background subtracted entire TR data trace (up to 300 ps) for the two samples, giving a clear mode at ∼ 0.1 THz in both cases, are plotted in Fig. 2(e) and (f), respectively. Next, we have scanned the time-resolved traces (with an interval of 100 ps), and performed the FFT in each successive time window. The resulting frequency vs. time profiles for the samples measured at different magnetic fields and pump fluences are summarized in Fig. 3. In all cases, a distinct mode is observed at ∼ 0.1 THz, which we assign to the nutation. Further details of the data analysis are discussed in the Supplemental Materials [29]. The nutation frequency shows a negligible dependence on the applied magnetic field and pump fluence. The TR sum signals are provided in the Supplemental Materials [29], where we do not see any mode at ∼ 0.1 THz. This is a strong criterion for interpreting our 0.1 THz driven oscillation as purely magnetically driven. An alternative interpretation for such high-frequency dynamics could be exchange-dominated perpendicular standing spin-wave (PSSW) modes across the film thickness. However, we can exclude these modes to be relevant here, since theoretically predicted PSSW modes in Py films of similar thickness appear at much higher frequencies. The calculated PSSW mode frequencies are ∼ 1 THz and ∼ 0.5 THz for 2.8 nm and 5 nm thick Py films, respectively. Additionally, the PSSW modes exhibit an inverse relationship with sample thickness, which is not observed here.

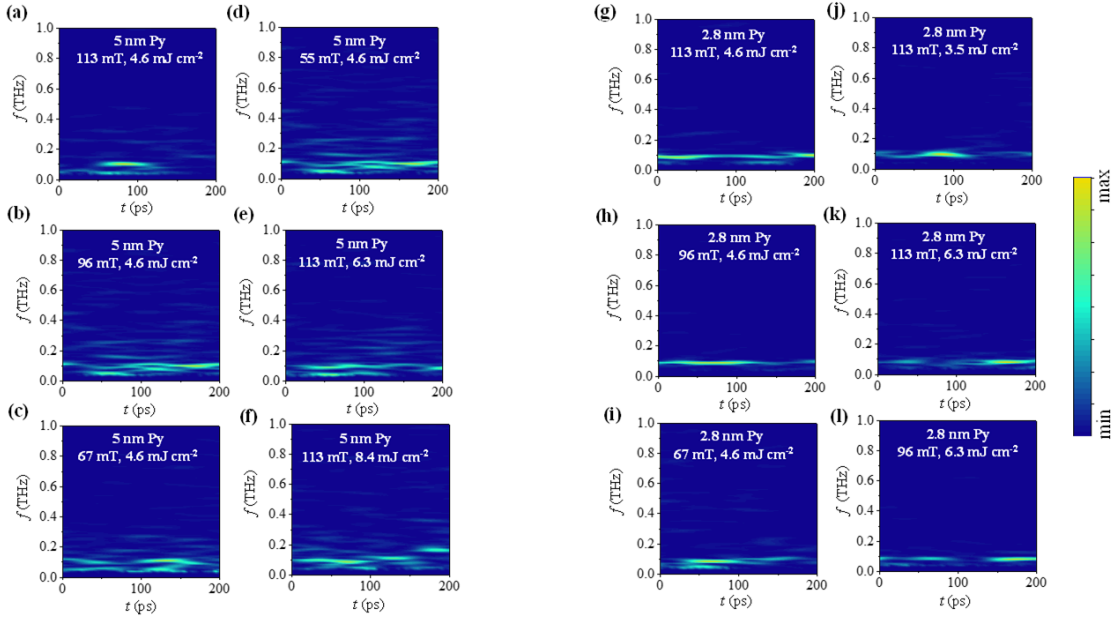

**FIG. 3.** Frequency vs. time plots obtained from the FFT of the background subtracted experimental time-resolved Kerr rotation signals for the 5 nm and 2.8 nm Py samples at different magnetic fields and pump fluences showing a clear mode at ~ 0.1 THz in each case. The sample thickness and values of $\mu_0 H_{ext}$, $F$ are shown in the respective plots.

### B. Inertial Landau-Lifshitz-Gilbert equation

For nearly 90 years, magnetizations dynamics has been described by the Landau-Lifshitz [LL-equation], which – with minor modifications by Gilbert – is now known as the Landau-Lifshitz-Gilbert (LLG) equation [1, 36],

$$\frac{d\vec{M}}{dt} = -\gamma \vec{M} \times \left[ \mu_0 \vec{H}_{eff} - \frac{\alpha}{\gamma M_S} \frac{d\vec{M}}{dt} \right] \tag{1}$$

where $\gamma$ is the gyromagnetic ratio, $M_S$ the saturation magnetization, $\alpha$ the Gilbert damping constant, and $\vec{H}_{eff}$ the effective field which includes contributions from an external magnetic field but also from exchange interactions or anisotropies. The first term on the right-hand side of equation (1) accounts for the precession of magnetization vector ($\vec{M}$) around $\vec{H}_{eff}$. The second term with a first-order time derivative of $\vec{M}$ is the Gilbert damping term [36], which occurs due to the transfer of energy and angular momentum of $\vec{M}$ to the environmental degrees of freedom and leads to a relaxation of $\vec{M}$ towards the direction of $\vec{H}_{eff}$. However, in deriving the LLG equation, the assumption is made that, analogous to the rigid body, only one component of the diagonal inertial tensor in the body-fixed system is finite [36]. Considering an inertial tensor of the form $I = \text{diag}(I_1, I_1, I_3)$ leads to an additional second-order time derivative term in the equation of motion that plays the role of inertia for the magnetization.

The resulting inertial Landau-Lifshitz-Gilbert (ILLG) equation reads [8, 37],

$$\frac{d\vec{M}}{dt} = -\gamma \vec{M} \times \left[ \mu_0 \vec{H}_{eff} - \frac{\alpha}{\gamma M_S} \frac{d\vec{M}}{dt} - \frac{\eta}{\gamma M_S} \frac{d^2 \vec{M}}{dt^2} \right] \tag{2}$$

The second order derivative term on the right-hand side of equation (2) gives rise to inertia and can lead to an additional oscillatory motion superimposed on top of the usual precession dynamics, known as nutation. The nutation parameter $\eta$ linearly depends on $I_1$ and is thus directly linked to the new structure of the inertial tensor.

The time scale of the nutation is defined by $\eta$ which is expected to be in the range of ps or even less, leading to oscillations of much higher frequency (sub-THz range) than those associated with spin precession (GHz range). Note, that the definition of $\eta$ in equation (2) is slightly different to former work [8]. Here it is a time scale that is independent of the damping constant $\alpha$. Different types of derivations of the inertial term include phenomenological arguments [9] as well first principles calculations [11] and a relativistic approach based on the Dirac equation [15, 18]. The determination of the value of the parameter $\eta$ is, however, an open question, as different studies indicate values ranging from a few femtoseconds to hundreds of picoseconds [8, 10, 12, 27, 37]. Assuming a certain value for the parameter $\eta$, however, the angular frequencies for precession and nutation (for dynamics close to equilibrium) in a magnetic field $\mu_0 H_{\text{ext}}$ can be approximated from the following formula [25, 37]:

$$\omega_{\text{nu}} = -\frac{\sqrt{1+4\gamma\eta\mu_0 H_{\text{ext}}}+1}{2\eta} \approx -\frac{1}{\eta} \qquad \omega_{\text{p}} = \frac{\sqrt{1+4\gamma\eta\mu_0 H_{\text{ext}}}-1}{2\eta} \approx \gamma\mu_0 H_{\text{ext}} \qquad (3)$$

Similar expressions, which reproduce the first term of the expansions, were also derived in Ref. [13]. Note the different sign of $\omega_{nu}$ and $\omega_p$, which indicates that the sense of rotation of nutation is opposite to that of precession. In our experiments, we observed a nutation frequency of ~ 0.1THz, so that for our samples, $\eta = 1/\omega_{nu}$ comes out to be ~ 1.6 ps, which is about 5 times larger than the corresponding value reported in [27]. The differences in the nutation frequencies are intriguing and need to be discussed. We observe the nutation frequency in the same order of magnitude as previously observed [27], and in the range predicted by theory [38, 39]. In our work, a different excitation scheme may have caused some modifications in the system properties, affecting $\eta$. We performed measurements after strong ultrafast demagnetization by femtosecond optical pulse (~1.5 eV), which resulted in the generation of hot electrons above the Fermi level. In contrast, Ref. 27 used much lower photon energy and pulse intensity of the driving THz field (~1 meV), resulting in the absence of hot electrons and subsequent quenching of the magnetization on picosecond time scales. Therefore, we can conclude that Ref. 27 measured nutation in close-to equilibrium regime, while our case involved a regime far-from-equilibrium, which could affect $\eta$ and the nutation frequency. A more detailed discussion can be found in the Supplemental Materials [29] (see also references [37-40] therein). Further investigation of this time-dependent behavior is needed more insight into the dynamics.

### C. Theoretical analysis:

For a deeper analysis of the dynamics of magnetization and angular momentum in the inertial regime we perform atomistic spin dynamics simulations based on the classical Heisenberg model and the ILLG equation (details are provided in the Supplemental Materials [29]). To be able to compare with our experiments, we use a permalloy model from [41], where the atomic magnetic moments are located on an fcc lattice with 20% iron atoms and 80% nickel atoms. We use the nutation parameter of $\eta = 1.6$ ps obtained from our experiments above and a magnetic field $\vec{B} = (0, 110, 30)^T$ mT with an OOP component as in the experiments. Then we solve the stochastic ILLG equation on an atomistic level as described in the Supplemental Materials [29] and calculate the time-dependent magnetization curve $M_z(t)$ shown in Fig. 4(a). After a thermal excitation with a heat pulse the system shows the precessional dynamics around the effective field with a period in the range of hundred picoseconds. Due to the nutation a second oscillation on the single picosecond time scale arises. This can be seen in the inset of Fig. 4(a), showing that the nutational mode has a period in the range of some picoseconds. The Fourier transform in Fig. 4(b) shows, that the nutation frequency is ~ 0.1 THz. With these parameters, the nutation frequency is very similar to the experimentally determined one from Fig. 3 and also matches the analytical expectation from equation (3). Note, however, that here the actual frequency and not the angular frequency is considered.

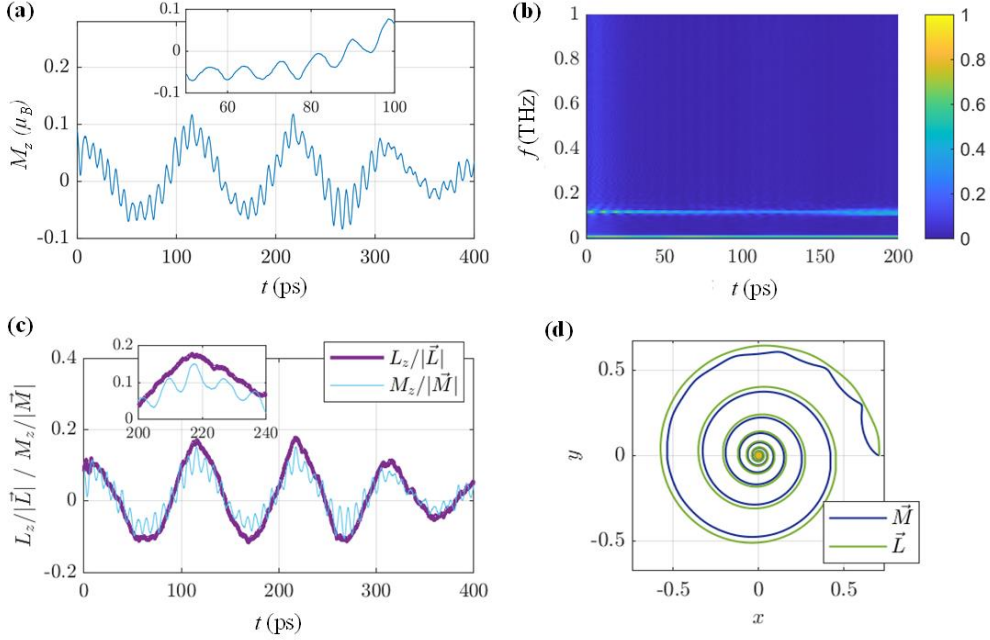

**FIG. 4.** (a) Time-dependent magnetization curve of the simulated Py model after an excitation with a rectangular ultrashort heat pulse. The magnetization shows both precession and nutation. (b) Frequency vs time plot of the spectral power density $P_z f^2$, unveiling the nutation frequency of ~ 0.1 THz and the precession frequency of ~ 0.01 THz. The spectral power density is multiplied with $f^2$ to reduce the thermal background noise. (c) Temporal dependence of the reduced magnetization and angular momentum of the out-of-plane component, showing that the magnetization nutates around the angular momentum (d) Sketch of the trajectories of magnetization and angular momentum following a sudden tilting of the effective field, which is then perpendicular to the image plane.

Furthermore, our simulations allow us to investigate the behavior of the angular momentum $\vec{L}$ of the system, which is – due to the inertia – no longer parallel to the magnetization [37]. The relation between the magnetization $\vec{M}$ and the corresponding angular momentum $\vec{L}$ (per volume) is given by [37],

$$\vec{L} = \frac{1}{\gamma}\vec{M} - \frac{\eta}{\gamma M_S}\vec{M} \times \frac{d}{dt}\vec{M} \tag{5}$$

This equation also follows from the new structure of the inertial tensor with a finite $I_1$ and describes a separation of the directions of the magnetization vector and the angular momentum vector. In case of vanishing nutation (i.e. $\eta = 0$) these two vectors are aligned. They are separated for a finite value of $\eta$ and a time-dependent magnetization. The ratio between the magnetization and the part of the angular momentum pointing in magnetization direction (the projection of $\vec{L}$ on $\vec{M}$) is still given by the gyromagnetic ratio.

The temporal behavior of the system's angular momentum is shown in Fig. 4(c). On long time scales the angular momentum and the magnetization show the same precessional dynamics in the frequency range of single GHz, which is identical to the case of vanishing nutation. We point out that the nutation mode is only reflected in the magnetization, since only the magnetic moments nutate around their associated angular momentum. On the contrary, the angular momentum just precesses around the external field. As our simulations demonstrate, the nutation leads to a separation of the magnetic moment from its angular momentum, not only on the atomistic scale but also on a macroscopic level, splitting up $\vec{L}$ and $\vec{M}$ for a system of interacting magnetic moments. We experimentally measure the dynamics of the magnetization ($\vec{M}$), but not the angular momentum ($\vec{L}$), because the magneto-optical (MO) effects are given by the (transient) response of the spin-polarized electronic band structure. Therefore, a direct experimental verification of equation (5) is not possible. The sudden emergence of the nutation, as triggered by the laser pulse, can be understood when bringing the ILLG into an explicit form [37] that resemble Newton's equation of motion:

$$\eta \frac{d^2}{dt^2}\vec{M} = -\frac{\gamma}{M_S}\vec{M} \times (\vec{M} \times \mu_0 \vec{H}_{\text{eff}}) - \alpha \frac{d}{dt}\vec{M} - \frac{1}{M_S}\vec{M} \times \frac{d}{dt}\vec{M} - \eta \frac{1}{M_S^2}\vec{M}\left(\frac{d}{dt}\vec{M}\right)^2 \tag{5}$$

Initially we consider a resting magnetization $\vec{M}$ parallel to the effective field $\vec{H}_{\text{eff}}$. A sudden tilting (here via a fs laser pulse) of the effective field away from the magnetization direction starts the dynamics of both magnetization and angular momentum. As the initial magnetization rests, $\frac{d}{dt}\vec{M} = 0$ applies and only the first term on the right side of equation (5) is non-zero. Thus, the magnetization accelerates towards the effective field, perpendicular to the precession direction. After this initial acceleration the other terms have to be considered, leading to the magnetization dynamics as sketched schematically in Fig. 4(d). For the angular momentum, however, the initial acceleration of $\vec{M}$ leads to an immediate velocity $\frac{d}{dt}\vec{L} = -\frac{\eta}{\gamma M_S}\vec{M} \times \frac{d^2}{dt^2}\vec{M} = -\vec{M} \times \mu_0\vec{H}_{\text{eff}}$ according to equation (4) and equation (5), pointing in precession direction. So, while the magnetization responds indeed inert and accelerates due to the nutation term in the ILLG, the angular momentum instantaneously starts to precess, which makes it inertia-free in contrast to the magnetization. Consequently, the magnetization does not follow the angular momentum instantaneously after the excitation. An important indicator of the dynamics of nutation is that it decays more rapidly than precession.

## III. CONCLUSION:

In conclusion, using the TR-MOKE technique and atomistic spin dynamics simulations, we show that in thin $Ni_{80}Fe_{20}$ films coherent magnetic nutation can arise from a strong non-equilibrium of the spin system after optical excitation leading to transient separation of magnetization from its angular momentum. These findings cannot be explained by the ubiquitous Landau-Lifshitz-Gilbert equation but agree with an extension of this equation of motion that contains a second-order time derivative describing inertia. The magnetic nutation is in the high-frequency regime (close to THz range) and superimposed on the precession dynamics (GHz range). Our results demonstrate that the magnetic nutation and the corresponding separation of the angular momentum from its magnetic moment is an intrinsic phenomenon occurring during the relaxation of an ultrafast non-equilibrium excitation of a magnet and does not only rely on a resonant excitation of the spin system [27, 28]. Hence, an optical excitation of inertial spin dynamics can enable and control the integration of different magnetic processes, ranging from demagnetization via nutation to precession in one device. This will have significant consequences in the field of spintronics. The discovery of Einstein and de Haas [42] has connected magnetic moment with angular momentum, and, hence, linked electro-dynamics with mechanics. Our findings show that nutation can separate effects that rely on magnetization dynamics from those that rest on properties of the associated angular momentum. While most effects in spintronics rely on magnetic or magneto-optic effects, the recently discovered ultrafast transfer of spin angular momentum into the lattice [43] rests on (spin plus mechanical) angular momentum conservation. Similarly, the recently established research into chiral phonons [44, 45] can connect mechanical with magnetic degrees of freedom [46, 47]. The separation of these quantities – magnetic moment and angular momentum – on ultrashort time scales will have profound implications for the understanding of ultrafast spin related physics.

## ACKNOWLEDGMENTS

Funding for this work was provided by the Deutsche Forschungsgemeinschaft (DFG, German Research Foundation) under Grant No. TRR 173-268565370, Spin+X (Projects No. B11 and B03), Grant No. 318592081, and Grant No. 425217212-SFB 1432.

# *Supplemental Materials*

### S1. Time-resolved MOKE experiments

The experimental setup is shown schematically in Fig. 1(a) of the main article. Spin dynamics are measured by a TR-MOKE setup based on a two-color, non-collinear, all-optical pump-probe technique. An amplified 1 kHz Ti:sapphire laser system (Astrella, Coherent; $\lambda = 800$ nm, repetition rate $\sim 1$ kHz, and pulse width of $\sim 35$ fs) forms the experimental base, which is used to pump the spin dynamics. The second harmonic ($\lambda = 400$ nm) of the amplifier pulses is sent through a mechanical delay line and used to probe of the dynamics. The probe is normally incident on the sample, while the pump is incident at an angle ($\sim 30°$ with respect to the surface normal). An external magnetic field ($H_{ext}$) is tilted at a small angle ($\varphi \sim 15°$) to the plane of the sample. This provides a finite out-of-plane demagnetization field, which is transiently modified by the pump pulse to induce both precession and nutation of the spins. For a completely in-plane configuration of the external (and thus effective) magnetic field, no optically induced precession (and nutation) is observed. To exclude any non-magnetic signal, we perform the measurements for two opposite directions of the sample magnetization and extract the pure magnetic response from the difference of the two resulting Kerr signals. All experiments are carried out at room temperature and under ambient conditions.

### S2. Time-resolved sum signals

The representative time-resolved (TR) sum signals up to 300 ps for the two $Ni_{80}Fe_{20}$ (Permalloy or Py) samples measured at $\mu_0 H = 113$ mT and pump fluence $F = 4.6$ mJ cm$^{-2}$ are shown in Fig. S1 (a) and (b) respectively. The zoomed views of the TR sum signals for small time windows (100 – 160 ps) are shown in the inset of Fig. S1(a) and (b). The fast Fourier transformed (FFT) power spectra obtained from the sum signals (Fig. S1 (c) and (d)) show no clear mode at $\sim 0.1$ THz. Figure S1 (e) and (f) shows the comparison between the FFT power spectra obtained from the difference (black) and sum (red) signals. It is clearly observed that the power of the sum signals is negligible as compared to that of the difference signals. The frequency vs. time plots obtained from the time-resolved sum signals for the two Py samples measured at different magnetic fields and pump fluences are shown in Fig. S2. The sum signal is generally dominated by purely optically induced transient artifacts and shows no mode at $\sim 0.1$ THz for either Py sample. Thus, this is a strong criterion for interpreting our 0.1 THz oscillation as purely magnetically driven and arising due to the nutation of magnetization.

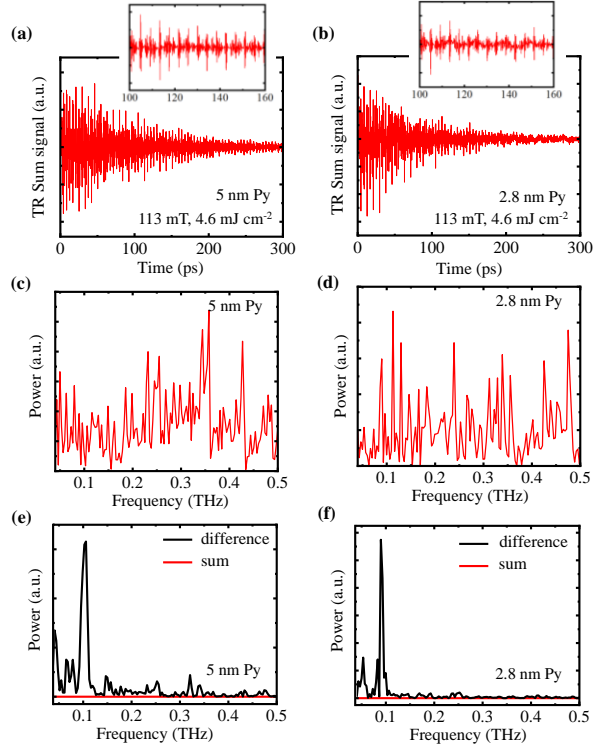

**FIG. S1. Demonstration of the absence of nutation in the sum signals:** Time-resolved sum signals measured up to 300 ps for (a) 5 nm and (b) 2.8 nm Py samples measured at $\mu_0 H_{ext}$ = 113 mT and $F$ = 4.6 mJ cm$^{-2}$. The zoomed views of the TR sum signals for small time windows (100 – 160 ps) are shown in the inset of Fig. S1(a) and (b). The corresponding FFT power spectra of (c) 5 nm and (d) 2.8 nm Py samples show no peak at ~ 0.1 THz. (e) – (f) The comparison between the FFT power spectra obtained from the difference (black) and sum (red) signals shows that the power of the sum signal is negligible compared to that of the difference signal. The $x$ axis is plotted from 30 GHz to 0.5 THz.

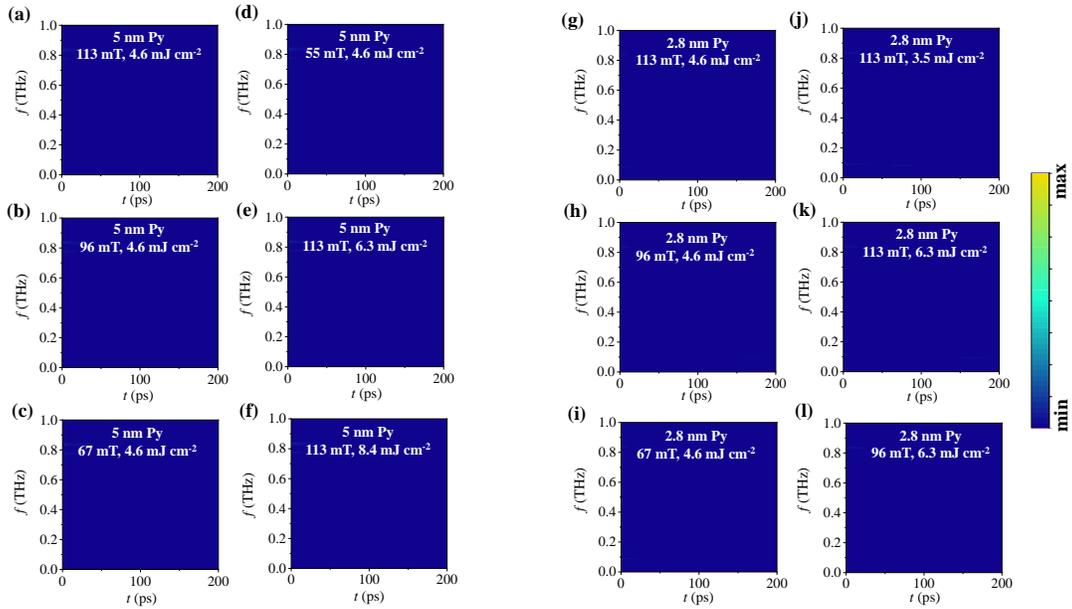

**FIG. S2. Demonstration of the absence of nutation in the sum signals:** Frequency vs. time plots obtained from the FFT of the TR sum signals for the 5 nm and 2.8 nm Py samples at different values of magnetic field and pump fluences. The sample thickness and the values of $\mu_0 H_{ext}$ and $F$ are shown in the respective plots. We do not observe any mode at ~ 0.1 THz in the non-magnetic sum signals.

## S3. Atomistic spin dynamics simulations

Permalloy (Py) is modeled as a doped fcc-lattice crystal consisting of 20% iron and 80% nickel atoms. We use a Heisenberg model to describe the interactions and energy contributions of the magnetic system. The magnetic moments are treated as classical, three-dimensional vectors $\vec{\mu}_i$ with a length of $2.637\ \mu_B$ for iron and $0.628\ \mu_B$ for nickel. $i$ indexes the lattice site, where $\mu_B$ is the Bohr magneton. In the following we will consider normalized and dimensionless magnetic moments $\vec{e}_i = \frac{\vec{\mu}_i}{\mu_i}$ with their corresponding magnitude $\mu_i$. Our Hamiltonian then reads

$$H\{\vec{e}_i\} = -\sum_{i,j} J_{ij}\vec{e}_i\vec{e}_j - d_z \sum_i (e_i^z)^2 - \vec{B}\sum_i \mu_i \vec{e}_i,$$

comprising an isotropic Heisenberg exchange, second order on-site anisotropy and the Zeeman term. The exchange parameters are taken from [41], considering the exchange up to a distance of 1.704 nm. To model the shape anisotropy $d_z$ is set to $-0.1$ meV, leading to a hard z-axis. The Zeeman term describes the coupling to the external field $\vec{B} = (0, 110, 30)^T$ mT.

To describe the dynamics, the explicit form of the inertial Landau-Lifshitz-Gilbert (ILLG) equation is used, which reads

$$\frac{d^2}{dt^2}\vec{e}_i = -\frac{\gamma}{\eta \mu_i}\vec{e}_i \times (\vec{e}_i \times \vec{H}_i) - \frac{\alpha}{\eta}\frac{d}{dt}\vec{e}_i - \frac{1}{\eta}\vec{e}_i \times \frac{d}{dt}\vec{e}_i - \vec{e}_i\left(\frac{d}{dt}\vec{e}_i\right)^2$$

and is equivalent to equation (2). The effective field

$$\vec{H}_i = -\frac{\partial H}{\partial \vec{e}_i} + \vec{\xi}_i,$$

describes the effect of the Hamiltonian on the dynamics of the magnetic moments. $\vec{\xi}_i$ is a Gaussian white noise with

$$\langle \vec{\xi}_i(t) \rangle = 0, \quad \langle \xi_{i\nu}(t)\xi_{j\eta}(t')\rangle = \frac{2\mu_i \alpha k_B T}{\gamma_i}\delta_{ij}\delta_{\nu\eta}\delta(t-t'),$$

taking into account the coupling to a heat bath with temperature $T$. $i$ and $j$ are lattice site indices, $\nu$ and $\eta$ represent the Cartesian directions. Due to the delta-functions on the right the noise is uncorrelated in both, space and time. On this basis the temporal evolution of a system of $N = 64^3$ spins with periodic boundary conditions is numerically computed via the stochastic Heun method. In addition, the angular momentum $\vec{L}_i$ for each atomic magnetic moment is calculated by means of [37],

$$\vec{L}_i = \frac{\mu_i}{\gamma}\vec{e}_i - \frac{\eta \mu_i}{\gamma}\vec{e}_i \times \frac{d}{dt}\vec{e}_i$$

The gyromagnetic ratio is set to the one of a free electron $\gamma = 1.76 \cdot 10^{11}\ \frac{1}{\text{Ts}}$, the damping constant is set to $\alpha = 0.02$ and the temperature after the excitation to $k_B T = 25$ meV. The excitation is modelled by means of a rectangular heat pulse in the beginning of the simulation with a width of 8 ps and a height of $k_B T = 400$ meV. Note, that the length of the heat pulse in our simulations has to be much larger than that of the laser pulse. Due to the inertia, the spin system reacts slower, so that even the demagnetization takes more time. This is a strong indication that the ILLG equation alone cannot describe the initial demagnetization phase completely but additional electronic processes also contribute to the demagnetization on femtosecond time scales.

The main output of the simulation is the system's magnetization

$$\vec{M}(t) = \frac{1}{N}\sum_i \frac{\mu_i}{\mu_B}\vec{e}_i(t),$$

which is here the averaged magnetic moment per lattice site and dimensionless. The system's angular momentum is given by

$$\vec{L}(t) = \frac{1}{N}\sum_i \vec{L}_i(t).$$

The spectral power density of the magnetization in the Fourier domain is computed via

$$P_\beta(f) = \left\langle \frac{|M_\beta^T(f)|^2}{T}\right\rangle$$

with the truncated Fourier transform

$$M_\beta^T(f) = \int_0^T (M_\beta(t) - \langle M_\beta \rangle) e^{-i2\pi f t}\, dt$$

and with $\beta \in \{x, y, z\}$.

## S4. Details of data analysis

### A. *Fourier transform analysis of the experimental data*

With the FFT spectra shown in Fig. 2(e) and (f) of the main article, we wanted to extract the frequency components associated with the nutation oscillations superimposed on the coherent precession dynamics for thin Permalloy (or Py) films. To achieve this, we first recorded the TR-MOKE data up to 300 ps with very high (100 fs) time step resolution (as shown in Fig. 2(a) and (b) of the main article for 5 nm and 2.8 nm Py samples, respectively, measured at $\mu_0 H_{ext}$ = 113 mT and $F$ = 4.6 mJ cm$^{-2}$, and Fig. S4(a) and (b) of the Supplemental Materials, measured at $\mu_0 H_{ext}$ = 96 mT and $F$ = 4.6 mJ cm$^{-2}$). We then subtracted the Ferromagnetic Resonance (FMR) background with the frequency obtained from the FFT of the precessional oscillations [32]. It is worth mentioning that the precessional data up to ~ 900 ps are measured with the same all-optical TR-MOKE setup with a much lower time step resolution (1 ps). The detailed analysis of the precessional dynamics is reported in our previous work [32]. Nevertheless, we have provided the precessional data up to ~ 900 ps in the section S5. of the Supplemental Materials. After subtracting the FMR background from the TR Kerr rotation data in Fig. 2(a) and (b), we find that in addition to the FMR background of physical origin, there may be a slow drift in the TR Kerr rotation due to instrumental relaxation. The latter can also give rise to a huge peak near frequency zero in the FFT spectra. We use a polynomial function (order 5) to further subtract this background and then obtain the background subtracted TR data. We then performed an FFT of this resulting background subtracted TR data using Origin software with a Hanning window. Fig. 2(e) and (f) correspond to the FFT of the background subtracted entire TR data trace up to 300 ps. The same treatment is applied to all measured TR-MOKE data taken for the two samples at different applied magnetic fields and pump fluences. In all cases we obtain an FFT peak at ~ 0.1 THz. In this way, we obtain the ~ 0.1 THz mode from the entire TR data trace.

To further confirm the ~ 0.1 THz mode, we analyzed our TR data in a different way. We wrote a LabView program that can 'scan' through the background subtracted TR data up to 300 ps with an interval of 100 ps and perform the FFT in each individual window successively. This program also uses the Hanning window to perform the FFT as used in the Origin software. In this way, this LabView program allows us to confirm the ~0.1 THz mode also in smaller, successive time windows, and we get the resulting frequency vs. time plots as shown in Fig. 3 of the main article.

### B. *Analysis of the theoretical data*

The frequency vs. time plot in Fig. 4(b) of the main article is performed as follows: For five simulation runs with a length of 3 ns and a temporal resolution of 4 fs the time-dependent magnetization $m_\beta(t) - \langle m_\beta \rangle$ is Fourier transformed with the FFT method at a time interval of 100 ps. The mean value is subtracted to get rid of the peak at frequency zero. Then, the absolute value squared of the Fourier transformed magnetization is taken, which is modulo a pre-factor of the spectral power density. This spectral power density is then averaged over the simulation runs. Apart from the precession and the nutation peaks the spectral power density shows a $\frac{1}{f^2}$-dependence, which is the Brownian thermal noise. To eliminate the noise, the spectral power density is multiplied by $f^2$, resulting in a quantity being constant apart from the precession on the nutation peaks. This is then plotted in Fig. 4(b), where the time axis shows the different starting points of the Fourier transformed time intervals, each 100 ps in length.

## S5. Time-resolved Kerr rotation data up to longer time delays

The background subtracted time-resolved Kerr rotation data up to longer time delays (~ 900 ps) and with lower time step resolution (1 ps) showing precessional oscillations for the two Py samples measured at two different magnetic field strengths and at pump fluence of $F = 4.6$ mJ cm$^{-2}$ are shown in Fig. S3. The black lines are the experimental data and the red lines are damped sinusoidal fits to the data points. These are data (up to longer time delays) corresponds to Fig. 2 (a) and (b) of the main article and Fig. S4 (a) and (b) of the supplemental materials. From the damped sinusoidal fit and also from the FFT (performed in the Hanning window) of these precessional oscillations, we can extract the precession (FMR) frequencies. A detailed analysis of these precessional oscillations is discussed in our previous report [32].

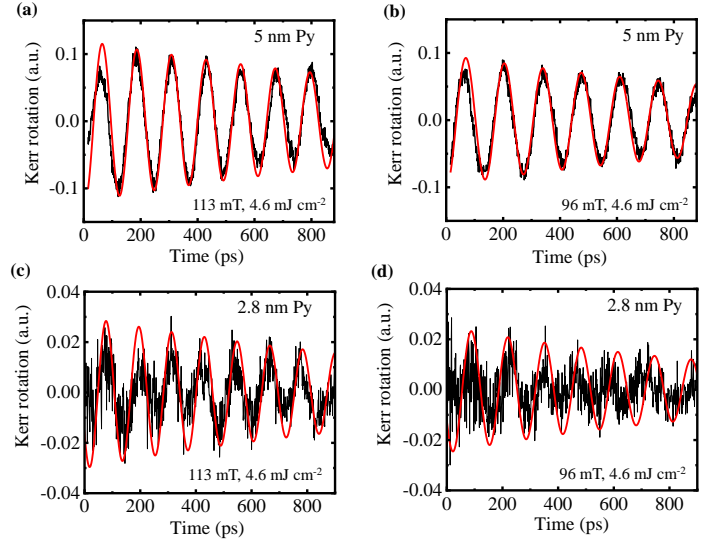

**FIG. S3.** (a) – (d) Background subtracted time-resolved Kerr rotation data up to longer time delays (~ 900 ps) showing precessional oscillations for the two Py samples measured at different magnetic fields and pump fluences. The sample thickness and the values of $\mu_0 H_{ext}$ and $F$ are mentioned in the respective figures. The black lines are the experimental data and the red lines are the damped sinusoidal fit to the data.

## S6. Experimental nutational data measured at a magnetic field $\mu_0 H_{ext}$ = 96 mT and pump fluence $F = 4.6$ mJ cm$^{-2}$

Experimental time-resolved Kerr rotation data measured up to 300 ps for 5 nm and 2.8 nm Py samples at $\mu_0 H_{ext}$ = 96 mT and $F = 4.6$ mJ cm$^{-2}$ are shown in Fig. S4. The black lines are the experimental data and the red lines are the FMR background. The zoomed views of the background subtracted TR data for a small time window (100 ps – 160 ps) showing nutational oscillations are depicted in Fig. S4(c) and (d). The black lines are the experimental data and the blue lines are the sinusoidal fit to the data points.

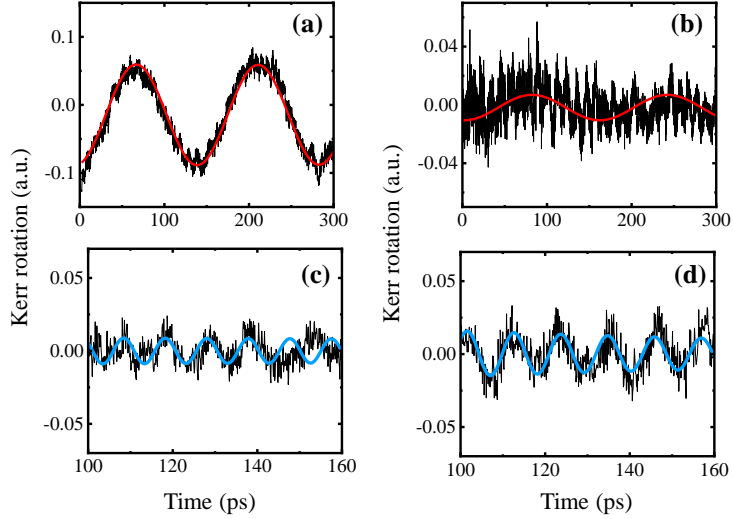

**FIG. S4:** Time-resolved Kerr rotation data measured up to 300 ps for (a) 5 nm and (b) 2.8 nm Py samples at $\mu_0 H_{ext} = 96$ mT and $F = 4.6$ mJ cm$^{-2}$. The black lines are the experimental data and the red lines represent the FMR background. Zoomed views of the background subtracted TR data of (c) 5 nm and (d) 2.8 nm Py samples for small time window (100 ps – 160 ps) showing nutational oscillations. The black lines are the experimental data and the blue lines are the sinusoidal fit to the data points.

## S7. Discussion of the discrepancy between the measured nutation frequency and that reported in *Nat. Phys. 17, 245 (2021)* [27]

In our experiments, we observe a nutation frequency of ~ 0.1THz, which yields a value of $\eta = 1/\omega_{nu}$ ~ 1.6 ps for our samples. This value is approximately 5 times larger than the corresponding value reported in Ref. [27] for the same sample ($Ni_{80}Fe_{20}$ or Permalloy). The differences in the measured nutation frequencies are intriguing and need to be discussed. Since the nutation frequency is mainly influenced by $\eta$, and the exact value and origin of $\eta$ is not fully known to the scientific community, the expected value of the nutation frequency is still a topic of intense debate [37]. However, we still observe the nutation frequency in the same order of magnitude as observed earlier [27], which is also in the range predicted by theory [38, 39]. The different excitation scheme in our work [in our case the system is non-resonantly excited by an ultrashort optical pulse and in the case of [27], the system was resonantly excited by a THz pulse] might have caused some modifications in the properties of the system, changing the value of $\eta$ and leading to a different nutation frequency in our case. We have measured the dynamics of the system after strong ultrafast demagnetization (highly non-equilibrium regime) caused by femtosecond optical pulse (~1.5 eV) excitation. During our nutation measurements, our system is still far from the equilibrium regime (still ≥ 30% of the magnetization is quenched and has not yet reached the equilibrium). This highly non-equilibrium regime is created by a high pump pulse intensity resulting in the generation of hot electrons above the Fermi level. In contrast, the previous report [27] uses a much lower photon energy of the driving THz field (~1 meV) and a much lower pulse intensity, which prevents the generation of such hot electrons and the subsequent quenching of the magnetization on picosecond time scales where the nutation is measured. Thus, we can conclude that in the previous report [27] nutation was measured in an equilibrium regime, whereas in our case it is measured in a far from equilibrium regime. This non-equilibrium state, initially caused by the generation of hot electrons above the Fermi level by strong femtosecond optical pulse excitation, could play an important role in modulating the nutation parameter $\eta$, leading to a different nutation frequency in our case.

Furthermore, our system differs in some aspects from Ref. [27]. In our work the $Ni_{80}Fe_{20}$ (Permalloy or Py) samples of two different thicknesses are epitaxially grown on MgO substrates by molecular beam epitaxy (MBE) technique. The thicknesses of our samples are 5 nm and 2.8 nm, which are much smaller than those of Ref. [27], where the sample was 15 nm thick. In Ref. [27], the samples were sputter deposited, and in our case, they are epitaxially grown on MgO substrate by MBE technique. Thus, our samples are different in terms of crystal growth, and such a difference in growth conditions may affect different parameters including $\eta$. The values of effective magnetization ($M_{eff}$) in our samples are much lower (~700 kA m$^{-1}$ for 5 nm Py and ~670 kA m$^{-1}$ for 2.8 nm Py) [32] as compared to those in Ref. [27] (~750 kA m$^{-1}$). Our samples are capped by 3 nm Al$_2$O$_3$ layer, while in Ref. [27], the Py samples were capped by 1.5 nm Al layers. It has been previously reported that different capping layers can affect the effective magnetization ($M_{eff}$) and effective damping ($\alpha_{eff}$) of the system [40], which in turn can change the nutation parameter $\eta$. Since our samples are much thinner as compared to the previous report [27], this may lead to some surface anisotropy and affect $\eta$, as also discussed in literature [38, 39]. All these factors could have caused a slight modification of $\eta$, resulting in a different nutation frequency in our case.